# Local Probing of Mesoscopic Physics of Ferroelectric Domain Walls

Vasudeva Rao Aravind,[1,2,*] A.N. Morozovska,[3] I. Grinberg,[4] S. Bhattacharyya,[1] Y. Li,[1]

S. Jesse,[5] S. Choudhury,[1] P. Wu,[1] K. Seal,[4] E.A. Eliseev,[6] S.V. Svechnikov,[3]

D. Lee (이동화),[7] S.R. Phillpot,[7] L.Q. Chen,[1] A.M. Rappe,[4] V. Gopalan,[1,†] and

S.V. Kalinin[5,‡]

[1] Materials Research Institute and Department of Materials Science and Engineering,

Pennsylvania State University, University Park, PA 16802

[2] Physics Department, Clarion University of Pennsylvania, Clarion, PA 16214

[3] V. Lashkarev Institute of Semiconductor Physics, National Academy of Science of Ukraine,

41, pr. Nauki, 03028 Kiev, Ukraine

[4] Department of Chemistry, University of Pennsylvania, Philadelphia, PA

[5] Center for Nanophase Materials Sciences, Oak Ridge National Laboratory,

Oak Ridge, TN 37831

[6] Institute for Problems of Materials Science, National Academy of Science of Ukraine, 3,

Krjijanovskogo, 03142 Kiev, Ukraine

[7] Department of Materials Science and Engineering,

University of Florida, Gainesville FL 32611

---

[*] This author was previously known by the name Aravind Vasudevarao
[†] vgopalan@psu.edu
[‡] sergei2@ornl.gov

Domain wall dynamics in ferroic materials underpins functionality of data storage and information technology devices. Using localized electric field of a scanning probe microscopy tip, we experimentally demonstrate a surprisingly rich range of polarization reversal behaviors in the vicinity of the initially flat 180° ferroelectric domain wall. The nucleation bias is found to increase by an order of magnitude from a 2D nucleus at the wall to 3D nucleus in the bulk. The wall is thus significantly ferroelectrically softer than the bulk. The wall profoundly affects switching on length scales of the order of micrometers. The mechanism of correlated switching is analyzed using analytical theory and phase-field modeling. The long-range effect is ascribed to wall bending under the influence of a tip bias well below the bulk nucleation field and placed many micrometers away from the wall. These studies provide an experimental link between the macroscopic and mesoscopic physics of domain walls and atomistic models of nucleation.

## I. Introduction

Dynamics of interfaces in materials and their interaction with microstructure and defects is the key element determining functionality of electrochemical systems,[1,2,3] ferroelectric and ferromagnetic materials, and phase transformations.[4] Interface dynamics controls the energy storage density in batteries and capacitors, switching speed, critical bias, and retention in ferroic[5] and phase change memories, and microstructure and properties of materials.[6] The role of interface behavior in materials science, energy and information technologies has stimulated an intensive effort on understanding the relationship between electronic, atomic, and mesoscopic structures and dynamic behavior of the interface.

Domain walls separating regions with opposite ferroelectric polarization are the prototypical example of interfaces in ferroic materials and have been extensively studied over the last 60 years.[7] The narrow width of the 180° wall necessitates the formation of the 2D nuclei as a rate-limiting step in wall motion and results in strong lattice and defect pinning.[8] Notably, similar motion mechanisms operate at phase transformation and solid-state reaction fronts and other high-energy interfaces. On the mesoscopic scale, wall-defect interactions give rise to a rich spectrum of dynamic behaviors[9,10] reflected in the complex self affine wall geometries observed down to ~10-30 nanometer length scales.[11,12]

The synergy between electron and scanning probe microscopies has allowed comprehensive understanding of the *static* domain wall structures at atomic and mesoscopic scales.[13,14,15] Switching of ferroelectric domains generated in the two limits of extremely large fields applied far away from the domain wall (i.e. bulk switching through nucleation) or smaller fields applied at the domain wall (i.e. field-induced domain wall motion) have been investigated in previous work and are likewise now well-understood. In the intermediate region, a number of observations,[16,17,18] including the correlated nucleation at the moving domain wall front,[19,20] suggest that the walls can strongly affect the properties of adjacent material due to long-range electrostatic and elastic fields. Nevertheless, fundamental questions, such as whether the nucleation energy of a 2D nucleus[21] on the wall can be measured directly, and especially the effect of the wall on the nucleation in the vicinity of the wall [22] have never been answered experimentally.

Here, we report on experimental studies of the nucleation behavior of ferroelectric domains using the spatially localized electric field of a biased scanning probe microscopy tip. This allows us to directly measure the intrinsic critical voltage for the formation of 2D nucleus at the wall[23,24] as well as to reveal the influence of the wall on the nucleation in the bulk. Surprisingly, we find that nuclei formed in the bulk interact with the domain wall even at extremely large micron-scale range, significantly lowering the barriers for domain nucleation. These finding have obvious implications for dynamics of polycrystalline ferroelectrics, and similar mechanisms can be operational in other systems with high-energy interfaces, including electrochemical systems and solid-solid transformations.

## II. Switching Spectroscopy PFM of a Ferroelectric Wall

Here we study *local* dynamic behavior of ferroelectric domain wall using spatially localized electric field of a biased scanning probe microscopy tip. The application of the *local* field to the wall results in the local wall displacement, and associated change of electromechanical response of the surface is detected as Piezoresponse Force Microscopy signal. This approach allows probing intrinsic (as opposed to extended defect-controlled) polarization dynamics, since the number of defects within the probing volume is necessarily small. The spatial extent and strength of the electric field acting on the wall can be controlled in a broad range by varying tip-wall separation and tip bias.

Near stoichiometric (NS) crystal of z-cut lithium niobate (LN), 900 nm thick was used for this study. Indium tin oxide (ITO) electrode was deposited on a +z surface by magnetron sputtering to establish bottom electrode. The NSLN sample was mounted with its −z surface upward on the 0.5-mm-thick congruent lithium niobate substrate using organic adhesive. Conductive silver paste was used to establish electrical contact with the bottom ITO electrode. To create the ferroelectric domain wall, the polarization in LN single crystal was reversed by the application of a high (44-66 V) bias pulse to the SPM tip, resulting in a macroscopic (~2 μm) domain of a characteristic hexagonal shape as shown in Fig. 1 (a). The effective tip radius was calibrated from the observed wall width and the bulk nucleation potential, as described in Section III.1.

To address nanoscale polarization switching dynamics in the presence of domain wall, we utilize Switching Spectroscopy Piezoresponse Force Microscopy (SS-PFM).[25] SSPFM is implemented on a commercial SPM system (Asylum MFP-3D) equipped with external data-acquisition electronics based on NI-6115 fast DAQ card to generate the probing signal and store local hysteresis loops and correlate them with surface topography. In SSPFM, the tip approaches the surface of the sample vertically with the deflection signal being used as the feedback, until the deflection set-point is achieved. Once the tip-surface contact at the predefined indentation force is established, the piezo motion is stopped and a hysteresis loop is acquired. During the acquisition of a hysteresis loop in SSPFM, the tip is fixed at a given location on the surface of the sample and the wave form $V_{tip} = V_{dc} + V_{ac} \cos \omega t$ is applied to the tip. Here, $V_{ac}$ is the amplitude of the PFM driving signal and the corresponding frequency is typically in the 200-500 kHz range. The probing signal $V_{dc}$ is the DC bias applied to the tip formed by the triangular wave (0.1 – 1 Hz) modulated by square wave (~100 Hz) to yield on-field and off-field responses. Application of sufficiently high DC bias results in the nucleation and subsequent growth of domains of opposite polarity below the tip, with a concurrent change of the PFM signal from *PR* (initial state) to *–PR*(switched state). The resulting *PR* dependence of DC bias contains information on domain nucleation and growth below the tip. In SSPFM, the hysteresis loops are acquired at each point in an $M \times N$ grid that is further analyzed to yield 2D maps of switching parameters such as work of switching, imprint, etc.

Here, the local electromechanical hysteresis loops are acquired over densely spaced (10 nm) grid of points (60x60 pixels) and analyzed to produce 2D maps of switching parameters. To ensure the reversibility of tip-induced wall displacement, the PFM images were acquired before and after the SS-PFM mapping. While the domain wall shifts on average, the length travelled (~1 pixel per line scan in the image and ~30 pixels total per image) is small compared to the total number of measurement points (3600). Thus, the wall dynamics is almost reversible. The measurements are performed as a function of bias window (i.e. maximal bias during the hysteresis loop acquisition) to decouple the bias and distance effects on wall dynamics.

The 3D data sets and 2D SSPFM maps contain the information on the domain nucleation in the presence of the wall. The averaged piezoelectric response image in Fig. 1 (b) shows dark and bright regions with no switching (that correspond to the original domains), and the region of intermediate contrast. The examination of the hysteresis loops illustrates that the loops are closed in the bright and dark regions, and are open in the region of intermediate contrast. The bright regions in Figure 1 (d) shows that in the vicinity of the domain wall the work of switching (WoS) (i.e. area under the loop) is non-zero and the hysteresis loops are open even below the threshold bias for bulk nucleation. The bright region is quite large, indicating that the presence of the domain wall has a strong influence on polarization dynamics even at a long range.

Analysis of the SSPFM data as a function of bias window (the maximum amplitude of dc bias applied to the probe) quantifies the dependence of the switching behavior as a function of nucleus distance from the domain wall. The dynamic evolution of the initial response, imprint (i.e. lateral shift of hysteresis loop along the voltage axis), and WoS are shown in Fig. 2. For small bias windows (< 3V), the hysteresis loops are closed, WoS is zero, and the initial response map is similar to the PFM image. For bias window > 3V, the intermediate contrast region in the initial response image and the white feature in the work of switching image emerge, indicating the onset of domain wall mobility. As the bias is increased, these features slowly extend into the region far away from the original domain wall. Surprisingly the boundary between the switching and non-switching regions extends as far as 1 μm for voltages of 15-25 V. While larger than the 3 V required for nucleation at the domain wall, are still significantly smaller than the large bias value (>28 V) for which the SS-PFM contrast disappears and bulk nucleation is observed.

In this intermediate voltage regime, the phase does not show ~180° hysteresis at all points. Note that if the amplitude shows a hysteresis behavior but the phase does not show a hysteresis behavior (no 180 degree switching), the work of switching will still show a high value (because WOS only reflects the mixed piezoresponse). To determine the bulk nucleation field, a point far from domain wall (at least about 5 microns) was chosen and bias voltage was applied incrementally, starting from low values ~5V. It was only at about 28 to 32V that switching behavior with 180º shift in the piezoresponse phase hysteresis was observed. So bulk nucleation was considered to be >28V at a frequency of 320 kHz.

Examination of the imprint image reveals an additional difference between the traditional switching through a 2D nucleus on the domain wall and the new nucleation mechanism responsible for the nucleation far away from the domain wall at intermediate voltages. The imprint images exhibit a complex structure, with imprint almost zero at the wall and forming strong maximum and minimum at the boundaries of the affected region. This behavior is indicative of the strong asymmetry of the hysteresis loop for tip positions to the left and to the right of the domain wall, as can be directly verified by the examination of the loop shape from individual locations [Fig. 3]. The dynamic regimes observed as a function of probe-wall separation and bias window are summarized in Fig. 3, delineating the regions of no switching, bulk switching, and wall-mediated switching.

Interestingly, the transition lines in Fig. 3 between no switching and asymmetric switching and between asymmetric and symmetric switching are can be correlated to the framework of standard theory of phase transitions. Defining a dynamic order parameter $\zeta = \int \{PR^+(V) - PR^-(V)\}dV$, i.e. area under the loop, it is clear that the transition between the no-switching and switching regime is second order (Figure 4). Similarly, the transition between wall-mediated and bulk switching regimes is first-order for order parameter defined as $\xi = \int \{PR^+(V) + PR^-(V)\}dV$, i.e. the average signal.

These experimental observations suggest two non-trivial observations regarding the mesoscopic physics of ferroelectric domain wall as explored by SSPFM. The observations of the minimal tip bias for the domain wall displacement suggest that the critical bias corresponding to the formation of 2D Miller-Winreich nucleus is measured directly. We also find that in addition to the standard nucleation mechanisms (formation of the 2D nuclei on the domain wall and bulk nucleation), an unexpectedly long-range interaction between the wall and the domain nucleus gives rise to a third, previously unexamined pathway for ferroelectric domain nucleation and switching manifested as an unexpectedly long-range (micron scale) effect of the wall on the domain nucleation bias and loop shape. The nucleation mechanism is analyzed in Section III using combination of first-principles, atomistic phase field, and analytical theory. The origins of long-range domain wall effect on nucleation bias are explored in Section IV using combination of analytical theory and phase-field modeling.

### III. 2D nucleation at domain walls

To get insight into the mechanism of 2D nucleation at the domain wall, we utilize the first-principles density functional theory to obtain estimates of domain wall energy and the height of the Peierls barrier. This data is combined with the atomistic phase-field model to yield the estimate of nucleation bias, and these estimates are further compared with Miller-Weinreich and Burtsev-Chervonobrodov semiclassical models for 2D nucleation.

### III.1. Tip parameters estimation

Comparison between the two extremes of bulk nucleation and nucleation at the wall allows us for the first time to directly evaluate the activation energy of the 2D nucleus. Bulk nucleation takes place only at the high values of bias (> 28 V). Here, the applied electric field destabilizes one of the possible polarization orientations, transforming the potential energy surface (PES) from the ferroelectric double well to a single well, corresponding to the intrinsic switching in the tip-induced field.[26,27] The energy and the electric field required to do this can be estimated from the Landau theory parameters for LiNbO$_3$.

The potential distribution induced by the probe, $V_e(x,y)$, was approximated as $V_e(x,y,d) \approx Vd/\sqrt{x^2 + y^2 + d^2}$, where $V$ is the applied bias and $d$ is the effective probe size.[28]. Then we identify the effective size of the tip $d = 86$ nm from bulk nucleation bias $V_c = 28$ V using $d = V_c\sqrt{27\beta\varepsilon_{11}\varepsilon_0/2\alpha^2}$, where $\alpha = -1.95 \cdot 10^9$ m/F and $\beta = 3.61 \cdot 10^9$ m$^5$/(C$^2$F) are expansion coefficients of LGD free energy on polarization powers for the second order uniaxial ferroelectrics, $\varepsilon_0$ is the universal dielectric constant, $\varepsilon_{11}=84$ is the component of the dielectric permittivity transverse to the polarization direction. (see Ref. [22] and Appendix A in the Supplementary Materials).

For tip positioned directly at the domain wall, the application of the bias results in wall bending with an associated change in the electromechanical response. Due to the presence of lattice pinning, the formation of a 2D Miller-Weinreich nucleus as the elementary step of wall motion requires a finite probe bias to be applied to the tip, and results in the opening of the hysteresis loop. This behavior is directly observed in Fig. 3, where the potential $V_i = 3$ V corresponds to wall unpinning from the lattice, or, equivalently, to formation of a nucleus at the wall. The effective tip size determined from the bulk nucleation threshold allows this wall-mediated process to be explored in Section IV.4.

### III.2. Domain wall energetics

We performed the calculations of the domain wall energy[29] for the domain wall lying at the cation plane and between the anion planes, using both density functional theory calculations (DFT) and atomic-level methods with the empirical potential by Jackson et al.[30]

DFT method is known to be more accurate than empirical method, but computationally expensive. DFT calculation yields a Y-wall energy of 160 mJ/m$^2$ when the center of Y-wall is sitting between two anion planes. For the Y-wall at the cation plane, the maximum domain wall energy of 310 mJ/m$^2$. Assuming that this is the maximum in energy, these calculations yield a Peierls barrier to wall motion of 150 mJ/m$^2$. The corresponding analysis using the empirical potential yield an energy of 230 mJ/m$^2$ with the Y-wall center lying between anion planes, and 485 mJ/m$^2$ with the center of Y-wall at the cation plane. Thus the (presumably less reliable) empirical study yields a Peierls barrier of 255 mJ/m$^2$, which is twice larger value than DFT study. The fit of polarization profile to hyperbolic tangent function derived from GLD theory estimates the domain wall width of 2.12Å. The lattice periodicity along the Y-wall is 2.58 Å. Thus, the ratio between domain wall width δ and the lattice periodicity d, δ/d of 0.82, potentially yielding a high Peierls potential.

### III.3. Atomic-level phase-field modeling

The quantitative description of the 2D nucleation process at the domain wall is achieved using recently developed diffuse nucleus model. Here, the polarization profile around the nucleus on the domain wall is given by a generalized form of the well known polarization profile for the 180 degree domain wall.

$$P_z(x,y,z) \approx 2P_s f^-(x,l_x,\delta_x)f^-(y,l_y,\delta_y)f^-(z,l_z,\delta_z) + P_z^{180}(x-l_x/2,y,z) \quad (1)$$

where $2f^\pm(a,b,g) = \tanh((2a+b)/g) \pm \tanh((2a-b)/g)$, $l_k$ corresponds to the length of the nucleus to the $k$ direction, and $\delta_k$ corresponds to the diffuseness to the $k$ direction. The $P_z^{180}(x-l_x/2,y,z)$ term corresponds to the polarization profile of the initial flat domain wall.

The free energy change $\Delta U$ of a nucleus due to the external field $E$ acting on the 180º domain wall is a sum of the local energy, gradient energy and field-polarization terms and is given by $\Delta U = \Delta U_v + \Delta U_i$, where

$$\Delta U_v = -E \int_{-\infty}^{\infty}\int_{-\infty}^{\infty}\int_{-\infty}^{\infty} dxdydz \left(P_z(x,y,z) - P_z^{180}(x,y,z)\right), \quad (2)$$

and

$$\Delta U_i = \int_{-\infty}^{\infty}\int_{-\infty}^{\infty}\int_{-\infty}^{\infty} dxdydz \left[\left(g_x\left(\frac{\partial P_z}{\partial x}\right)^2 + g_y\left(\frac{\partial P_z}{\partial y}\right)^2 + g_z\left(\frac{\partial P_z}{\partial z}\right)^2 + U_{loc}(P_z)\right) - \left(g_x\left(\frac{\partial P_z^{180}}{\partial x}\right)^2 + U_{loc}(P_z^{180})\right)\right] \quad (3)$$

The subscripts $v$ and $i$ refer to *volume* and *interface*, respectively. The local contribution is $U_{loc}(p) = A_{loc}\left(1 - (p/p_s)^2\right)^2$, where $A_{loc}$ is the ferroelectric well depth at 0 K and $g_x$ and $g_z$ parameterize the energy cost of longitudinal and transverse polarization changes. The contribution of the depolarization energy is ignored as it is negligible for a small nucleus. Therefore, parameters $A_{loc}(T)$, $g_x$, $g_z$, and $P_s(T)$ are necessary to evaluate the energy of the critical nucleus.

The temperature dependence of $A_{loc}(T)$ is obtained from the DFT $A_{loc}$ at 0K and the ratio of the experimental $P_s$ at finite temperature and at 0 K. For LiNbO$_3$, the 0 K DFT polarization is ~0.8 C/m$^2$ [Ref. [31]]. The $g_x$ parameter is proportional to the square root of $(\sigma_{100}/P_s)^2/A_{loc}$, where $\sigma_w$ is the 180º domain wall energy and $P_s$ is the 0 K polarization (see Eqn. (16) in Suppl. Mat. For Ref. [24]). To evaluate $g_x$, we used the DFT value of the LiNbO$_3$ 180 degree domain wall energy $\sigma_w$=0.159 J/m$^2$ and $A_{loc}$ =0.25 eV [Ref. [32]]. This results in $g_x$ value of 9.9×10$^{-12}$ m$^3$/F. To estimate the $g_z$ gradient parameter, we use the LiNbO$_3$ $g_x$ value and the ratio between $g_z$ and $g_x$ ($g_z$ / $g_x$ = 1.7) previously found for PbTiO$_3$. The $g_z$ parameter is then equal to 1.68 ×10$^{-11}$ m$^3$/F .

Using the electric field strength (5×10$^7$ V/m) at the domain wall corresponding to the experimentally observed voltage of 3 V and solving numerically for the polarization profile with minimum energy at different $l_k$ and $\delta_k$, we obtain $l_y$ ~ 12 Å, $l_z$ ~ 20 Å, $\delta_y$ ~ 4 Å, $\delta_z$ ~ 6 Å and a critical nucleus energy of 0.17 eV or about 7 $k_BT$, sufficient for almost instantaneous nucleation. Although there are several sources of uncertainty in the calculation of the critical nucleus energy, such as the variation in the values of $A_{loc}$ and room temperature $P_s$ (e.g. see Ref. [33]), these will not change the value of the critical nucleus energy by a large enough amount to make the nucleation time longer or comparable to the experimental ~ms timescale. This suggests that the activation barrier for 2D nucleation at the wall is controlled by the intrinsic nucleation process.[24]

**III.4. Semiclassical models for 2D nucleation**

The atomistic model in Section III.3 estimates the activation barrier for 2D domain nucleation as 0.17 eV, if the depolarization field effects are ignored. In this section, we analyze the effects of depolarization contributions on the 2D nucleation using extensions of Miller-Weinreich (MW) and Burtsev-Chervonobrodov (BC) models. We note that MW considered the lattice discreteness in very rough model and do not take the possibility of the wall to bent into account as well as the wall is regarded infinitely thin. In contrast to MW smooth Burtsev-Chervonobrodov (BC) approach considered much more realistic model with continuous lattice potential and diffuse domain walls, at that the nucleus shape and domain wall width are calculated self-consistently.

For the MW model, the activation energy for 2D rigid nucleus formation in the electric field of a biased PFM tip averaged over the nucleus volume is given by

$$F_a^{MW}(\sigma_W, V, x_0) = \frac{8}{3\sqrt{3}} \sqrt{\ln\left(\frac{\langle\sigma_W\rangle\theta}{2cP_S d^2 V}\right) \frac{(c\langle\sigma_W\rangle)^3}{\pi\varepsilon_0\varepsilon_{11}} \frac{\theta}{d^2 V}}, \quad (4)$$

where $\sigma_W$ is the domain wall energy, parameter $\theta = \gamma\left(\sqrt{d^2 + x_0^2}\right)^3$ originated from the averaging of the tip electric field

$$E_3 = \frac{V(d + z/\gamma)d}{\gamma\left((d + z/\gamma)^2 + \rho^2\right)^{3/2}} \quad (5)$$

over the nucleus volume ($\rho = \sqrt{x^2 + y^2}$), $c$ is the lattice constant, $\gamma = \sqrt{\varepsilon_{33}/\varepsilon_{11}}$ is the dielectric anisotropy factor, $x_0$ is the distance between the tip apex and the wall (see Appendix B in the Supplementary materials for details). The LiNbO$_3$ materials parameters are $c$ =0.5 nm, $P_S$=0.75 C/m$^2$, $\varepsilon_{11} = 84$, $\varepsilon_{33} = 30$. The domain wall energy $\langle\sigma_W\rangle = \sigma_{min} + \delta\sigma/2$ was calculated using density functional theory as $\sigma_W(x) \approx \sigma_{min} + \delta\sigma \sin^2(\pi(x - x_0)/c)$, where $\sigma_{min} = 0.160$ J/m$^2$, periodic lattice potential $\delta\sigma = 0.150$ J/m$^2$. [34] The activation barrier calculated using Eqn. (1) for applied voltage of 3 V and $d = 86$ nm is much more than 50 $k_B T$, i.e. corresponds to observation time $t = t_0 \exp(-F_a/k_B T) \sim 10^8$ s at phonon times $t_0 = 10^{-12}$ s, which is too high to account for reasonable experimental time. We estimate that according to MW model an applied voltage of 16-21 V would be required to unpin the domain wall from the lattice at $x_0$=0 (see Table 1 in Appendix B in the Suppl. Mat.).

Using BC approach we obtained the barrier directly at the wall ($x_0$=0):

$$F_a^{BC}(\sigma_W, V, x_0 = 0) = \sqrt{\ln\left(\frac{\gamma d\sqrt{\sigma_{min}\delta\sigma}}{2cP_S V}\right) \frac{(c\sqrt{\sigma_{min}\delta\sigma})^3}{4\pi\varepsilon_0\varepsilon_{11}} \frac{\gamma d}{V}}. \quad (6)$$

Let us underline the following distinctions between MW and expression (6):

(1) Replacement of $\langle\sigma_W\rangle = \sigma_{min} + \delta\sigma/2$ with $\sqrt{\sigma_{min}\delta\sigma}$, where $\sigma_{min}$ is the minimal value of potential and $\delta\sigma$ is the modulation depth.

(2) Due to the domain wall diffuseness, a sub-critical nucleus has a very smooth shape and factor $16\sqrt{3}/9$ disappears.

Eq.(2) yielding nucleation potential of 3.6 V for observation time ~10 ms (corresponding to barrier 25 $k_B T$), and 10.5 V for instant nucleation (corresponding to barrier of 1 $k_B T$). Note that these estimates are very close to that of the atomistic model, with primary uncertainly related to numerical values of LNO parameters and the contribution of depolarization field of the nucleus, and are fully consistent with experimental observations.

## IV. Long-range mesoscopic dynamics at the ferroelectric wall

The examination of the diagram in Fig. 3 illustrates the presence of the long-range interactions in the system exhibited as long-range effects of preexisting domain wall on nucleation bias and hysteresis loop shape. The nucleation bias is reduced by ~10% compared to bulk values at distances as large as ~2-3 μm, which is ~30 larger then tip radius estimated from either spatial resolution or bulk nucleation bias. To get insight into origins of this behavior, we perform the extensive phase field modeling of switching process, and develop an analytical long-range interaction model.

### IV.1. Phase field modeling of long-range interaction effects

The mesoscale mechanism of polarization switching in LiNbO$_3$ under PFM tip is modeled using phase field approach. In this, the ferroelectric domain is described by the spatial distribution of the spontaneous polarization vector $\vec{P}(\mathbf{x})$. The temporal evolution of polarization $\vec{P}(\mathbf{x})$ is obtained by solving the time dependent Ginzburg-Landau equation:

$$\frac{\partial P_i(\mathbf{x},t)}{\partial t} = -L\frac{\delta F}{\delta P_i(\mathbf{x},t)} \quad (i=1,2,3), \tag{7}$$

where $L$ is the kinetic coefficient associated with domain wall mobility and $F$ is the free energy functional. The free energy functional includes bulk, domain wall, elastic and electrostatic energies as $F = \int_V (f_{bulk} + f_{grad} + f_{elas} + f_{elec})dV$. For LNO, the bulk free energy $f_{bulk}$ is described by the Landau polynomial expansion as follows:

$$f_{bulk} = \alpha_1 P_3^2 + \alpha_{11} P_3^4 + \alpha_2 (P_1^2 + P_2^2), \tag{8}$$

where $\alpha_1 = -1.0 \times 10^9$ C$^{-2}$m$^2$N, $\alpha_{11} = 0.9025 \times 10^9$ C$^{-4}$m$^6$N, and $\alpha_2 = 0.9725 \times 10^9$ C$^{-2}$m$^2$N at room temperature. The gradient energy density $f_{grad}$ is non-zero at the domain walls and is described by

$$f_{grad} = \frac{1}{2}G_{11}\left(P_{1,1}^2 + P_{1,2}^2 + P_{2,1}^2 + P_{2,2}^2 + P_{3,1}^2 + P_{3,2}^2\right) + \frac{1}{2}G_{22}\left(P_{1,3}^2 + P_{2,3}^2 + P_{3,3}^2\right), \tag{9}$$

where $P_{i,j} = \frac{\partial P_i}{\partial x_j}$, $G_{11} = G_{22} = 0.4 G_0$, $G_0 = \alpha_0 (\Delta x)^2$, $\alpha_0 = -\alpha_1$, and $\Delta x$ is the grid size of the simulation box. The elastic energy $f_{elas}$ arises due to piezoelectric coupling between electrostatic fields and polarization and is given by

$$f_{elas} = \frac{1}{2}C_{ijkl}\left(\varepsilon_{ij} - \varepsilon_{ij}^0\right)\left(\varepsilon_{kl} - \varepsilon_{kl}^0\right), \tag{10}$$

where $C_{ijkl}$ is the elastic stiffness tensor, $\varepsilon_{ij}$ is the total strain, $\varepsilon_{ij}^0 = Q_{ijkl} P_k P_l$, with $Q_{ijkl}$ representing the electrostrictive coefficients. The non-zero electric stiffness and electrostrictive coefficient in the Voigt's notation are $C_{11} = 1.99 \times 10^{11}$ Nm$^{-2}$, $C_{12} = 0.55 \times 10^{11}$ Nm$^-$

$^2$, $Q_{11}$=0.016 C$^{-2}$m$^4$, $Q_{12}$=-0.003 C$^{-2}$m$^4$, $Q_{44}$=0.019 C$^{-2}$m$^4$. Finally, the electrostatic energy density $f_{elec}$ is $f_{elec} = -\frac{1}{2}E_i(\omega_0 \kappa E_i + P_i)$, where $E_i = -\partial \phi / \partial x_i$ is the electric field, $\kappa$ is the relative permittivity, $\omega_0$ = 8.85x10$^{-12}$Fm$^{-1}$ is the dielectric permittivity of vacuum.

Eq.(7) is solved numerically using semi-implicit Fourier spectral method on a 128$\Delta x$ x 128$\Delta x$ x 64$\Delta x$ domain with periodic boundary conditions along $x_1$ and $x_2$ axes. The film thickness $h_f$ = 56$\Delta x$. The critical bulk nucleation potential is obtained by gradually increasing the potential $\phi_0$ at the tip in steps of 0.05 V until the new domain is observed.

The switching diagram of the tip bias voltage required for switching (i.e. for observing open PFM loops) as a function of the distance from the wall calculated using phase field modeling is shown in Figure 6. The circles in the figure are estimated using phase field simulations. The open circles represent the open loops and the closed loops are indicated by filled circles. The distances and the bias window voltage values are calibrated using experimental data. Also shown in dotted line is an approximate fit to the experimental data points from Figure 3. A series of example polarizations hysteresis loops at different tip positions for a fixed bias voltage of 16 volts is shown in Figure 6 (b). Note the excellent agreement between the experimental and theoretical switching diagrams. Here, we focus our comparison on two specific aspects of this diagram, namely (1) the switching threshold bias at the domain wall versus away from the wall, and (2) the long range influence of the wall up to several micrometers.

Interestingly, the phase-field modeling suggests the presence of threshold field for domain wall motion, despite the fact that the lattice-level pinning is not included explicitly. The threshold bias using phase-field modeling is finite and the magnitude of this bias depends on the number of time steps for which the system was relaxed under a bias field. In general for shorter period of relaxation, kinetic effects are still dominant and the bias threshold at the wall is higher. For example, for ~100 time steps, the bias threshold is ~3 volts, close to experimental results. For long relaxation times, the system approaches a steady state threshold of ~0.3 volts. The same value of steady state threshold bias was also observed for a simulation of the wall system under a uniform electric field instead of a biased tip, and likely represents the effects of the spatial discretization step in the phase field model. Overall, this suggests that experimentally observed 3V threshold corresponds to the formation of 2D nucleus controlled by the lattice periodicity effects.

The long range influence of the wall can be understood by tracking the domain wall evolution with time around a hysteresis loop, as shown in Figure 7. For tip biases below the bulk nucleation bias of 28 volts, the switching mechanism is predominantly governed by the attraction and repulsion of the wall due to the nearby tip. An opening in the polarization loop is observed when the wall bends and crosses past the region under the tip, and retraces back under a reverse bias. The asymmetry in the polarization loop arises from this fact, in that the loop opening occurs only for the bias that bends the wall towards the tip.

**IV.3. Analytical studies**

To understand the origins of long-range wall-tip interactions, we analyze the mesoscopic mechanism of polarization switching in the presence of an initially flat 180°-domain wall and in the absence of the lattice pinning. The dynamics of the polarization field, $P_3$, is described by the Landau-Ginzburg-Devonshire-Khalatnikov relaxation equation:

$$-\tau \frac{d}{dt} P_3 = \alpha P_3 + \beta P_3^3 - \eta \left( \frac{\partial^2 P_3}{\partial z^2} + \frac{\partial^2 P_3}{\partial x^2} + \frac{\partial^2 P_3}{\partial y^2} \right) - E_3, \tag{11}$$

where τ is the Khalatnikov coefficient, α<0 in ferroelectric phase, the gradient term $\eta > 0$, expansion coefficient $\beta > 0$ for the second order phase transitions considered hereinafter,. The polarization boundary conditions are $P_3 - \lambda \partial P_3 / \partial z = 0$ at the surface.

Electric field $E_3(x,y,z) = -\partial \varphi / \partial z$, where the electrostatic potential distribution, $\varphi(\mathbf{r})$, in ferroelectric is coupled with polarization as

$$\varepsilon_0 \varepsilon_{33}^b \frac{\partial^2 \varphi}{\partial z^2} + \varepsilon_0 \varepsilon_{11} \left( \frac{\partial^2 \varphi}{\partial x^2} + \frac{\partial^2 \varphi}{\partial y^2} \right) = \frac{\partial P_3}{\partial z}, \qquad (12)$$

where $\varepsilon_{33}^b \leq 10$ is the dielectric permittivity of background state[35] and $\varepsilon_0$ is the universal dielectric constant. $\varepsilon_{11}$ and $\varepsilon_{33} \gg \varepsilon_{33}^b$ are dielectric permittivity values perpendicular and along polar axis z. The potential distribution induced by the probe yields boundary conditions $\varphi(\rho, z=0) = V(t) d / \sqrt{\rho^2 + d^2}$, where V is the applied bias and d is the effective probe size, hereinafter $\rho = \sqrt{x^2 + y^2}$.

Allowing for the principle of the electric field superposition, the electric field that satisfy Eq.(5) is the sum $E_3(x,y,z) = E_3^e(x,y,z) + E_3^d(x,y,z)$, where $E_3^e(x,y,z)$, is the probe field inside the sample and $E_3^d(x,y,z)$ is the depolarization field created by the curved domain wall [36]. Expressions for the fields are listed in Appendix A of the Supplement. Note that ferroelectric cubic nonlinearity ($\sim \beta P_3^3$) and the order parameter spatial dispersion (polarization gradient) determine only the short-range correlation effects between the domain nucleus and the wall, which dominate at distances $|x_0| \leq d$. However the Coulombic electric field is mainly responsible for the long-range interaction between the slightly curved domain walls and the probe-induced domain nucleus located even far enough (i.e. at $|x_0| \gg d$) from the walls. Actually, the power decay of Coulombic field possibly results in correlated switching at distances more than hundred nanometers.

Using direct variational method for polarization redistribution $P_3(x,y,z,t) \approx P_0(x) + P_V(t)f(x,y,z)$, where $P_0(x) = P_S \tanh((x-x_0)/2L_\perp)$ is the initial flat domain wall profile positioned at $x=x_0$ (the correlation length is $L_\perp = \sqrt{-\eta/2\alpha}$, and the spontaneous polarization is $P_S^2 = -\alpha/\beta$ ). The coordinate-dependent part

$$f(\rho,z) \approx \frac{\sqrt{\varepsilon_{11}\varepsilon_0/(-2\alpha)}d^2}{(L_\perp d + d^2 + \rho^2)\sqrt{d^2+\rho^2}} \text{ satisfy the linearized Eq.(4).}$$

The parameter $P_V$ serves as effective variational parameter describing domain geometry, and allows reducing complex problem of domain dynamics in the non-uniform field to an algebraic equation obtained after the substitution of $P_3(x,y,z,t)$ into the LGD free energy functional, integration and minimization on $P_V$.

Thus we derived that in the presence of lattice pinning of viscous friction type, the amplitude $P_V$ should be found from the equation of Landau-Khalatnikov type:

$$\tau \frac{d}{dt}P_V + w_1 P_V + w_2(x_0)P_V^2 + w_3 P_V^3 = V(t) \quad , \tag{13}$$

where constants $w_i$ describe tip geometry and materials properties as:

$$w_1 \approx 1, \quad w_2(x_0) \approx \frac{-3\beta P_S x_0}{\sqrt{(L_\perp+d)^2 + x_0^2}} \frac{\sqrt{-2\alpha\varepsilon_{11}\varepsilon_0}}{4\alpha^2(L_\perp+d)}, \quad w_3 \approx \frac{\beta\varepsilon_{11}\varepsilon_0}{4\alpha^2(L_\perp+d)^2} \tag{14}$$

Eqs. (4)-(7) provide comprehensive description of polarization dynamics in the vicinity of the wall and in the bulk.

Introducing the new parameter $P = P_V \dfrac{\sqrt{-2\alpha\varepsilon_{11}\varepsilon_0}}{-2\alpha(L_\perp + d)} - P_S$, one can rewrite the static Eq.(6) as

$$\alpha P(6n(x_0) - 5) + \beta P^3 + 3\beta P^2 P_S (1 - n(x_0)) = \dfrac{\sqrt{-2\alpha\varepsilon_{11}\varepsilon_0}\, V}{L_\perp + d} - 3\beta P_S^3 (1 - n(x_0)) \qquad (15)$$

Let us underline that the terms determined by the function $n(x_0) = x_0 \big/ \sqrt{(L_\perp + d)^2 + x_0^2}$ originated from nonlinear (cubic) interactions of the probe field and the stray depolarization field with initially flat domain wall. As the result wall curvature or domain nucleation appears.

It is seen from Eq.(8) that far from wall ($x_0 \gg d$) it reduces to the usual symmetric ferroelectric hysteresis $\alpha P + \beta P^3 = V\sqrt{-2\alpha\varepsilon_{11}\varepsilon_0}\big/(L_\perp + d)$. Near the wall ($x_0 \ll d$) Eq.(S.7) gives equation $-5\alpha P + \beta P^3 + 3\beta P^2 P_S = \dfrac{\sqrt{-2\alpha\varepsilon_{11}\varepsilon_0}\, V}{L_\perp + d} - 3\beta P_S^3$ that reveals no hysteresis because of negative $\alpha$.

Static thermodynamic coercive biases $V_c^\pm(x_0)$ are determined from $dV/dP_V = 0$. The expected behavior of the hysteresis loops as a function of tip surface separation is illustrated in Fig. 5. Directly at the wall, the loop is closed and the local response originates from the bias-induced bending of the domain wall. The bistability is possible for $x_0^2 \geq 2(L_\perp + d)^2$. The loop width, determined as the difference of coercive biases $\Delta V_c = (V_c^+ - V_c^-)/2$, monotonically increases with the distance $x$. Far from wall ($x_0 \gg d$) corresponding coercive biases are symmetric, $V_c^\infty = \pm 2\alpha P_S (L_\perp + d)\big/\sqrt{-54\alpha\varepsilon_{11}\varepsilon_0}$. The inclusion of viscous friction leads to the loop broadening and smearing far from the wall, while near the wall the minor loop opening is observed (compare solid and dotted curves). Note that the qualitative evolution of hysteresis loops in Fig. 5 is highly reminiscent of the experimental data in Fig. 3.

**V. Summary**

To summarize, ferroelectric domain walls, long believed to be the simplest example of *static* topological defect in ferroic materials, are found to exhibit an unexpectedly rich panoply of nanoscale switching behaviors due to interplay between wall bowing and bulk-like nucleation. The effective nucleation bias is found to increase by an order of magnitude from a 2D nucleus at the wall to 3D nucleus in the bulk. The effect of the wall is extremely long range with significantly lower nucleation bias even for tip-wall separation in the μm range. This is due to the compensation of depolarization field of nascent domain by wall bowing. Notably, the nucleation bias at the wall (3 V) allows a direct measurement of the nucleation energy for the 2D nucleus, which is found to be well below that predicted by rigid ferroelectric (Miller Weinreich) models (~16-21 V), but is in an reasonable agreement with the smooth lattice potential models (Burtsev-Chervonobrodov) (~3-7 V) and in excellent agreement recently developed diffuse nucleus model.

Our studies open a pathway to detailed atomistic understanding of domain wall dynamic in ferroic materials, including wall-defect interactions (pinning), structure and behavior of the walls with coupled order parameters, and dynamic effects such as nucleation in front of the moving wall. In these, the biased probe representing local charged defect of controlled strength. These studies become increasingly important given the rapidly growing role of ferroelectrics and multiferroics in information and energy storage technologies.

Finally, we expect that the fundamental mechanisms explored in this work – namely the lowering of the potential barrier to the nucleation of a new phase induced by the presence of (mobile) interface due to screening of long-range electrostatic and elastic fields – will be applicable to a broad range of electrochemical and solid-solid phase transformations.

**Acknowledgements**

Research supported in part (SVK, SJ, KS) by Division of Scientific User Facilities, US DOE. AV and VG acknowledge the CNMS user program. VG acknowledges NSF-DMR-0820404, DMR-0908718, and DMR-0602986. DL and SRP acknowledge NSF DMR-0602986. ANM and SVS acknowledge joint Russian-Ukrainian grant NASU N 17-Ukr_a (RFBR N 08-02-90434). VG thanks K. Kitamura and K. Terabe for providing the thin crystal. Research sponsored by Ministry of Science and Education (UU30/004) of Ukrainian and National Science Foundation (Materials World Network, DMR-0908718). The authors acknowledge multiple discussions with J.F. Scott, P. Paruch, A. Levanyuk, A. Gruverman, and A.K. Tagantsev and thank them for invaluable advice.

**Figure Captions**

**Figure 1.** (a) Mixed Piezoresponse force microscopy images of the domain structure before and the SSPFM scan with $\pm 5.0$V voltage window. Note that while domain wall moved during the experiment (dotted line within the square), the high veracity of SS-PFM PFM map indicates that no significant wall rearrangement was happening during single pixel or scan line acquisition. The piezoresponse (b), imprint (c) and work of switching (d) SS-PFM map in the domain wall region. (e-h) Hysteresis loops from selected locations.

**Figure 2**. Evolution of the wall dynamics as a function of bias window. Shown are (a) piezoresponse, (b) work of switching, and (c) imprint SS-PFM maps. The images are corrected (aliasing) to compensate for wall creep during measurements.

**Figure 3.** Switching phase diagram showing polarization dynamics as a function of bias window and tip-wall separation. Shown are the regions of no switching (light blue regions), wall-mediated switching with asymmetric loops (yellow regions) and symmetric loops (orange region near $x=0$), and bulk nucleation (orange region). Red lines correspond to first-order phase transitions across which the switching loops change discontinuously from open to closed. Blue lines correspond to second order phase transitions from symmetric to asymmetric switching loops. Blue dotted line marks a continuous transition between wall-mediated and bulk responses. The threshold bias for polarization reversal at the wall is $V^*_i$, and in the bulk is $V_b$. Shown below are experimental PFM hysteresis loops or a bias window of $\pm 14$ volts and at $x=$ -220nm, -140nm, 0nm, +80nm, and +120nm.

**Figure 4.** Order parameters profiles across 180º domain wall in lithium niobate, obtained using SSPFM. (a) Work of Switching and (b) Average PFM signal, (c) Imprint and (d) Vertical offset. The legend shown to the right applies to (a), (b), (c) and (d).

**Figure 5.** Energies of Y-walls at the cation plane and between anion planes. The equilibrium position is determined as the center between two anion planes. The curves are guides to the eye.

**Figure 6.** (a) Phase-field prediction of switching phase diagram. The open and closed circles, calculated using phase field simulations, represent the open and closed loops, respectively. The broken red line is an approximate fit to the experimental data from figure 3 and the solid black line is an approximate fit to the phase-field predictions. The phase-field limit for nucleation bias at the wall is ~0.3 volts for 1000 steps relaxation (as compared to 0 in ideal model). The *x*- and *y*-axes in phase-field were calibrated using experimental values. (b) Five representative polarization hysteresis loops at steady state at different tip positions (-1080nm, -120nm, 0nm, +120nm, +1080nm) from the wall at a fixed tip bias voltage of 16 V.

**Figure 7.** Phase-field modeling results: (a) An asymmetric polarization hysteresis loop at a fixed tip bias of 16 V when the tip is positioned on the negative domain (-$P_s$) away from the wall. (b) Eight representative sections showing the evolution of the domain wall with time around the hysteresis loop. The sections correspond to the points marked with red circles and labeled 1-8 in the hysteresis loop on the left. The attraction and repulsion of the wall due to the nearby tip results in asymmetric switching.

**Figure 8.** Map of the switching regimes (upper row) and corresponding ferroelectric polarization hysteresis loop shape (insets a-e) depending on the wall-tip separation, $x_0$, calculated from the analytical theory. Dotted curve (and loops a-e) is plotted for the static case $\tau=0$, solid curves (and loops a-e) correspond to the kinetic case with $\tau \neq 0$. The reversible wall bending occurs at $0<x_0<d$, correlated nucleation occurs at $d < x_0 < 10d$, and symmetric bulk nucleation starts at $x_0 \gg 10d$. Curves correspond to the different relaxation coefficients $\tau=0$, $10^{-8}$, $10^{-7}$, $10^{-6}$ SI units. Material parameters for LiNbO$_3$ are $L_\perp = \sqrt{-\eta/2\alpha}$ ~0.5 nm, extrapolation length $\lambda \gg L_\perp$, $\eta=10^{-9}$ m$^3$/F, $\varepsilon_{11}$=84, $\varepsilon_{33}$=30, $\alpha=-2\cdot10^9$ SI units, $\beta = 3.61\cdot10^9$ m$^5$/(C$^2$F), $P_S$=0.75 C/m$^2$, frequency $\omega=2\pi\cdot10^4$ rad sec$^{-1}$.

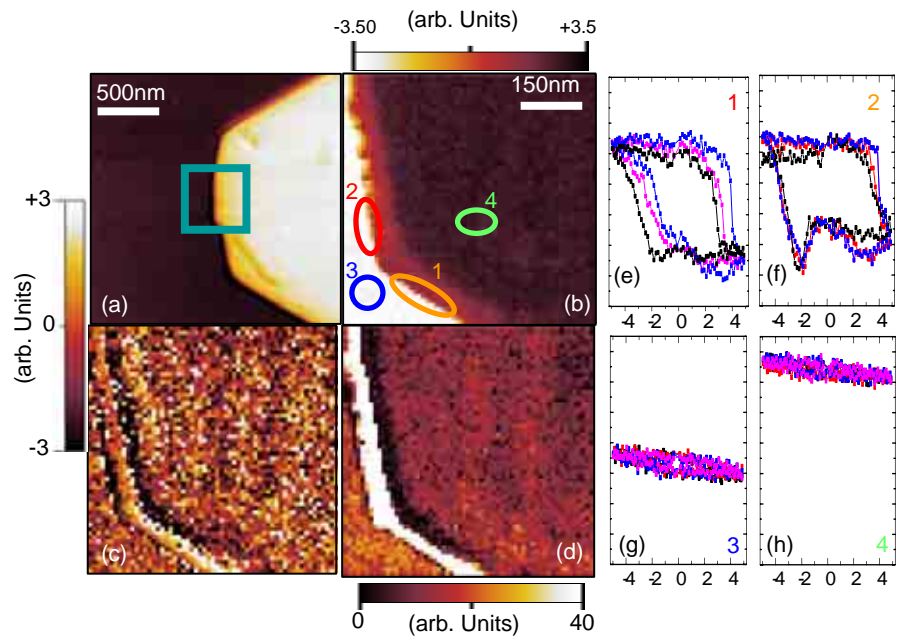

Figure 1

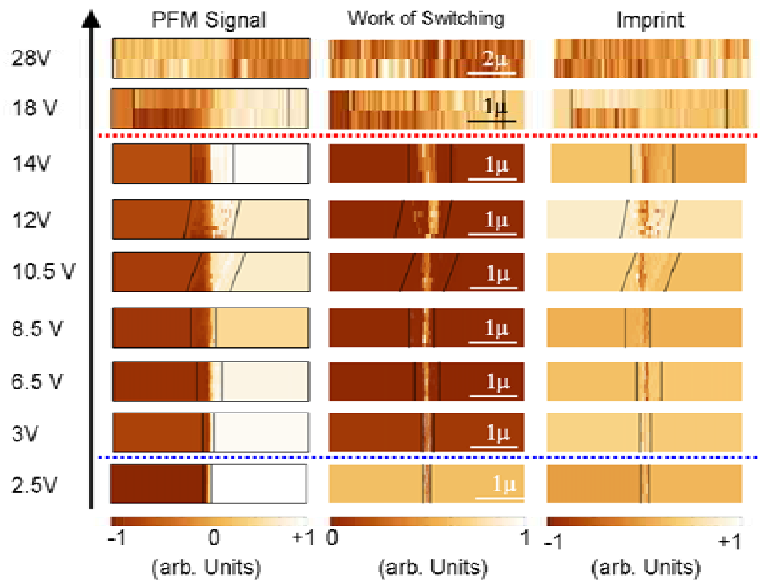

Figure 2

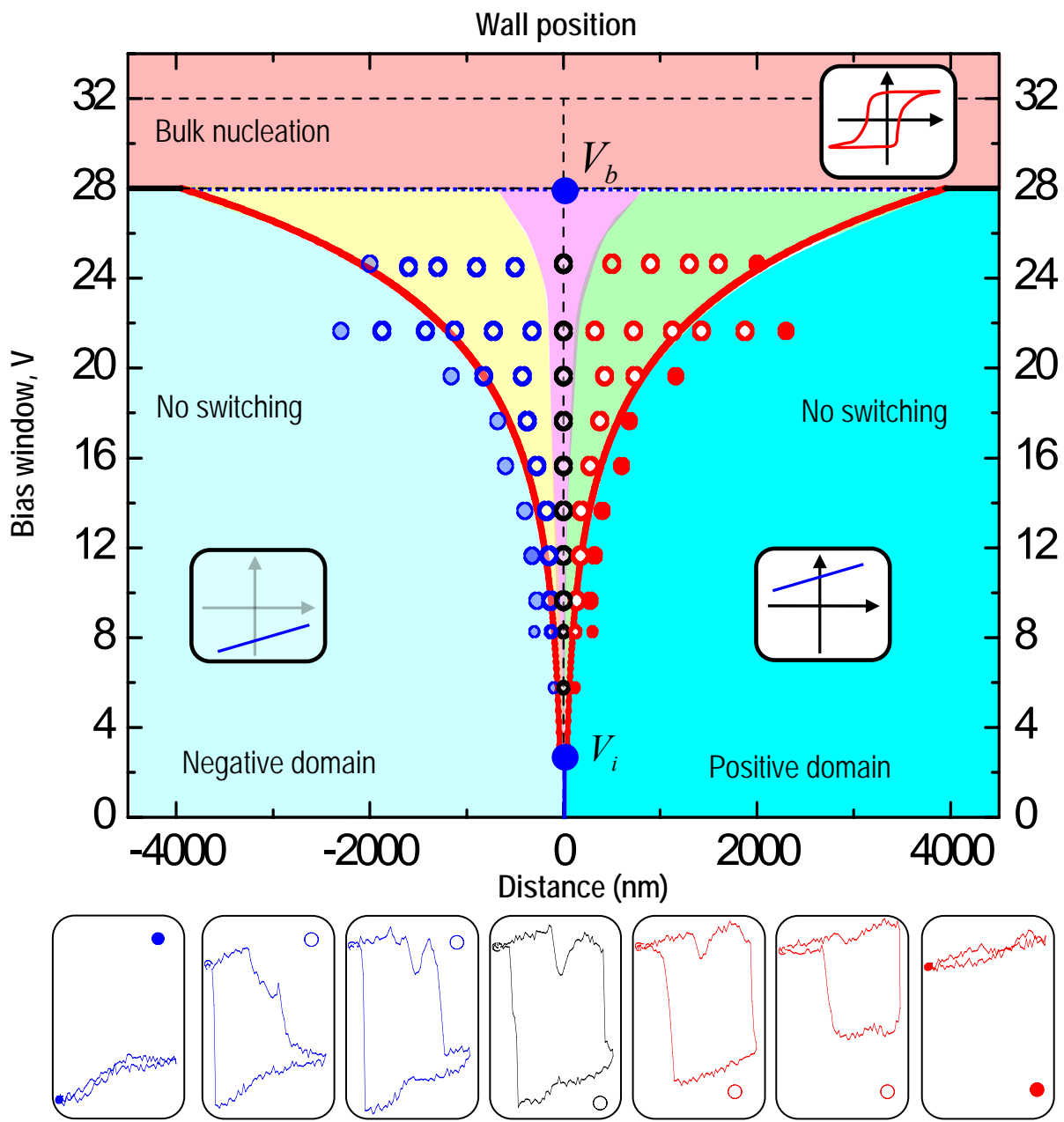

Figure 3

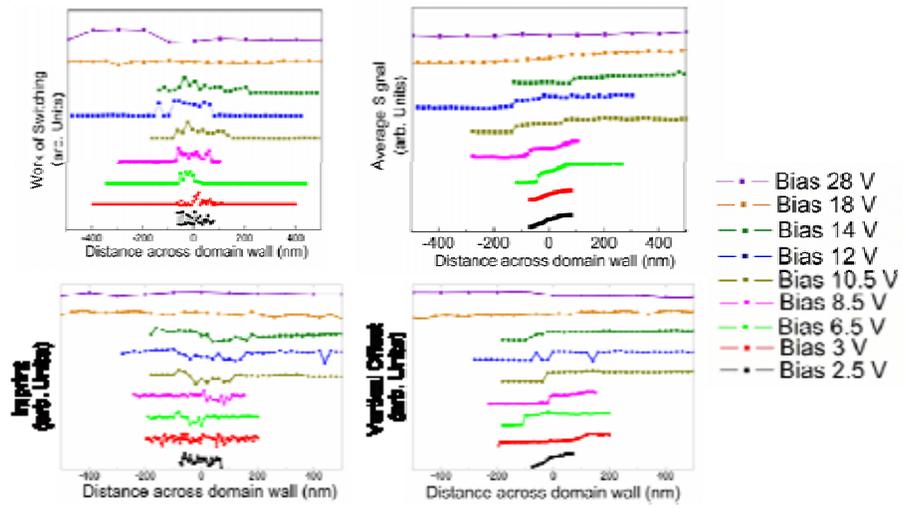

Figure 4

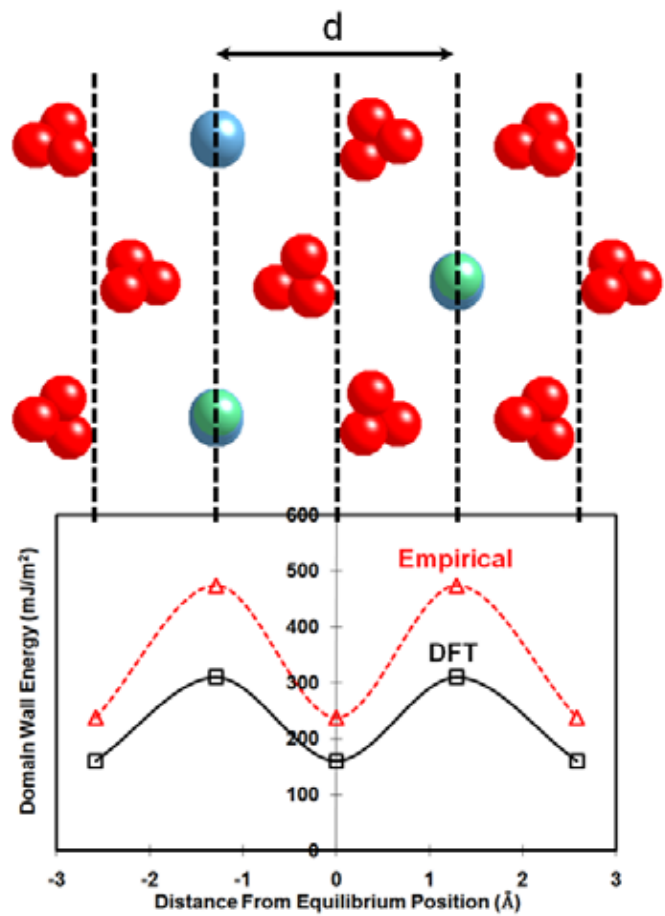

Figure 5

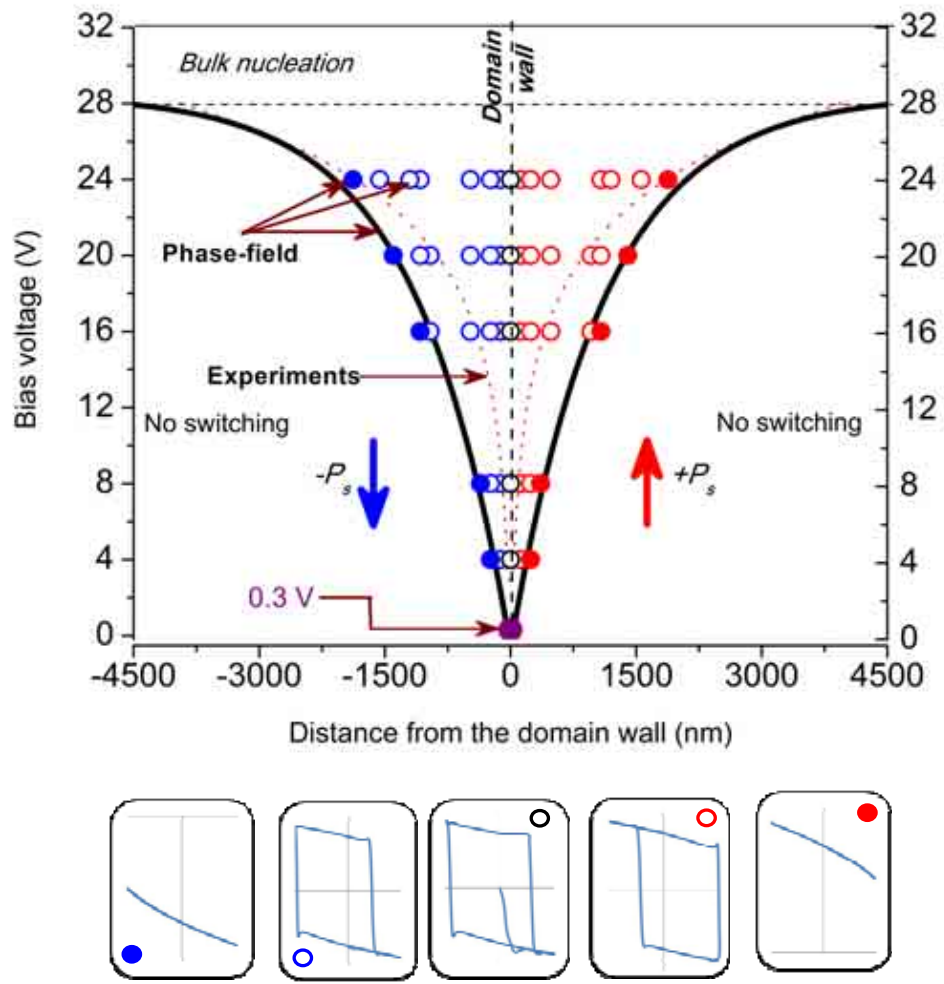

Figure 6

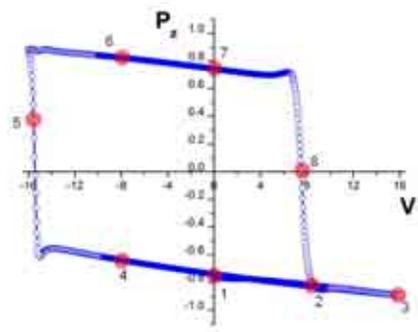 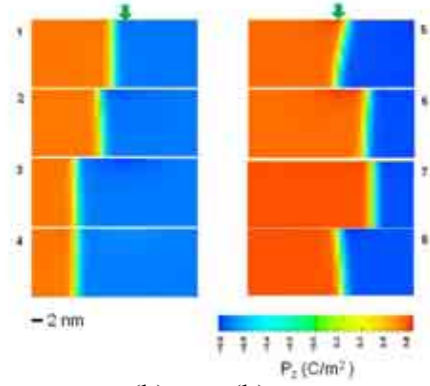

(a)            (b)    (b)

Figure 7

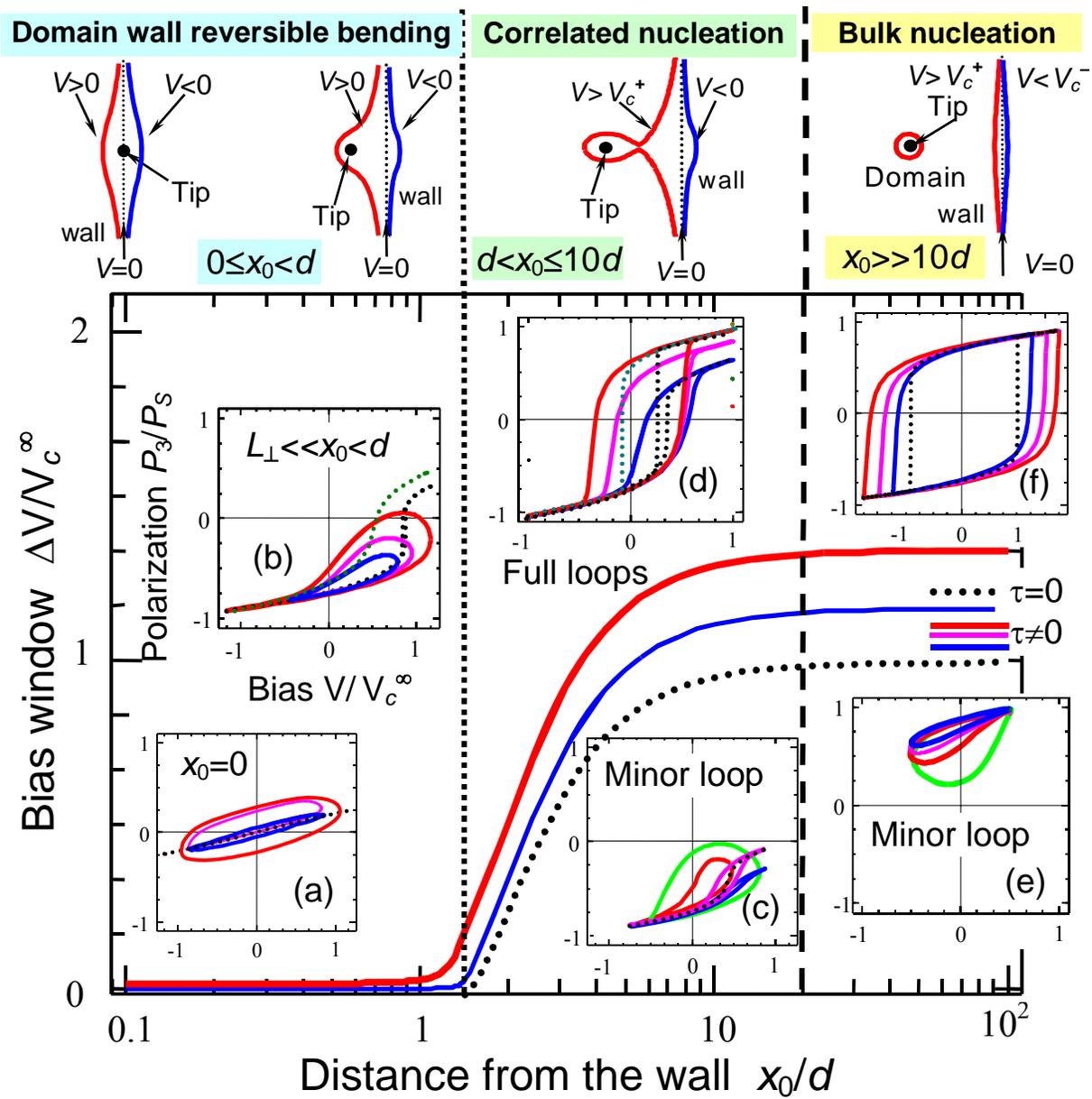

Figure 8

## Supplementary Materials

## Appendix A. Calculations within Landau-Ginzburg-Devonshire approach

### I. Basic equations

Landau-Ginzburg-Devonshire free energy for the uniaxial ferroelectric is

$$G(P, E_3^e) = \int_{-\infty}^{\infty} dx \int_{-\infty}^{\infty} dy \left( \int_0^h dz \left( \frac{\alpha}{2} P^2 + \frac{\beta}{4} P^4 + \frac{\xi}{2}\left(\frac{\partial P}{\partial z}\right)^2 + \frac{\eta}{2}(\nabla_\perp P)^2 - P\left(E_3^e + \frac{E_3^d}{2}\right)\right) + \frac{\xi}{2\lambda_1} P^2(z=0) + \frac{\xi}{2\lambda_2} P^2(z=h) \right) \quad (A.1)$$

where $\alpha < 0$ and $\beta > 0$ are expansion coefficients of LGD free energy on polarization powers for the second order ferroelectrics, electric field $E_3 = E_3^e + E_3^d$ is the sum of external and depolarization fields. Corresponding LGD-equation:

$$-\tau \frac{d}{dt} P_3 = \alpha P_3 + \beta P_3^3 - \xi \frac{\partial^2 P_3}{\partial z^2} - \eta\left(\frac{\partial^2 P_3}{\partial x^2} + \frac{\partial^2 P_3}{\partial y^2}\right) - E_3(x,y,z). \quad (A.2)$$

$\tau$ in kinetic coefficient. The electric field $E_3(x,y,z) = -\partial \varphi/\partial z$ can be expressed via electrostatic potential $\varphi(\mathbf{r})$.

The electrostatic potential distribution, $\varphi(\mathbf{r})$, in ferroelectric obeys the equation

$$\varepsilon_{33}^b \frac{\partial^2 \varphi}{\partial z^2} + \varepsilon_{11}\left(\frac{\partial^2 \varphi}{\partial x^2} + \frac{\partial^2 \varphi}{\partial y^2}\right) = \frac{1}{\varepsilon_0} \frac{\partial P_3}{\partial z}, \quad (A.3)$$

where $\varepsilon_{33}^b$ is the dielectric permittivity of background state[35] and $\varepsilon_0$ is the universal dielectric constant. The potential distribution induced by the probe yields boundary conditions $\varphi(x,y,z=0) = V_e(x,y)$. In the effective point charge approximation the distribution $V_e(x,y)$ can be approximated as $V_e(x,y,d) \approx V d/\sqrt{x^2 + y^2 + d^2}$, where $V$ is the applied bias and $d$ is the effective probe size.[28]

For more realistic modeling of the tip shape the summation over the image charges positions $d_i$ (or integration over the line charge in order to account for the conic part of the probe tip, see e.g. Refs. [37, 38, 39]) should be performed, namely

$$V_e(x,y) = \frac{V\left(\sum_i \dfrac{1}{\sqrt{x^2+y^2+d_i^2}} + \dfrac{1}{\ln(\mathrm{ctg}^2\theta/2)}\dfrac{2\varepsilon_e}{\varepsilon_e+\kappa}\ln\left(\dfrac{L+\Delta L+\sqrt{(L+\Delta L)^2+x^2+y^2}}{\Delta L+\sqrt{\Delta L^2+x^2+y^2}}\right)\right)}{\sum_i d_i^{-1} + \dfrac{2\varepsilon_e}{\varepsilon_e+\kappa}\dfrac{\ln(1+L/\Delta L)}{\ln(\mathrm{ctg}^2\theta/2)}}.$$

(A.4)

Hereinafter $\kappa = \sqrt{\varepsilon_{33}\varepsilon_{11}}$ is the effective dielectric constant, $\varepsilon_e$ is the ambient dielectric constant. The conical part potential is modeled by the linear charge of length $L$ with a constant charge density $\lambda_L = 4\pi\varepsilon_0\varepsilon_e V/\ln(\mathrm{ctg}^2\theta/2)$, where $\theta$ is the cone apex angle. The distance between the linear charge and the ferroelectric surface is $\Delta L$.

Allowing for the principle of the electric field superposition and linear electrostatic equations below we could consider the single-charge component $V_e(x,y,d)$ and the perform the integration/averaging in the final results. The corresponding Fourier representation on transverse coordinates $\{x,y\}$ of electric field normal component $\tilde{E}_3(\mathbf{k},z) = -\partial\tilde{\varphi}/\partial z$ is the sum of external ($e$) and depolarization ($d$) fields:

$$\tilde{E}_3(\mathbf{k},z) = \tilde{E}_3^e(V_e,\mathbf{k},z) + \tilde{E}_3^d(P_3,\mathbf{k},z), \qquad (A.4a)$$

$$\tilde{E}_3^e(\mathbf{k},z) = \tilde{V}_e(\mathbf{k})\frac{\cosh(k(h-z)/\gamma_b)}{\sinh(k h/\gamma_b)}\frac{k}{\gamma_b}, \qquad (A.4b)$$

$$\tilde{E}_3^d(P_3,\mathbf{k},z) = \begin{pmatrix} \int_0^z dz'\dfrac{\tilde{P}_3(\mathbf{k},z')}{\varepsilon_0\varepsilon_{33}^b}\cosh(k z'/\gamma_b)\dfrac{\cosh(k(h-z)/\gamma_b)}{\sinh(k h/\gamma_b)}\dfrac{k}{\gamma_b} + \\ \int_z^h dz'\dfrac{\tilde{P}_3(\mathbf{k},z')}{\varepsilon_0\varepsilon_{33}^b}\cosh(k(h-z')/\gamma_b)\dfrac{\cosh(k z/\gamma_b)}{\sinh(k h/\gamma_b)}\dfrac{k}{\gamma_b} - \dfrac{\tilde{P}_3(\mathbf{k},z)}{\varepsilon_0\varepsilon_{33}^b} \end{pmatrix} \qquad (A.4c)$$

Here $\gamma_b = \sqrt{\varepsilon_{33}^b/\varepsilon_{11}}$ is the "bare" dielectric anisotropy factor, $\mathbf{k} = \{k_1, k_2\}$ is a spatial wave-vector, its absolute value $k = \sqrt{k_1^2 + k_2^2}$. For typical ferroelectric material parameters the inequality $2\varepsilon_0 \varepsilon_{33}^b |\alpha| \ll 1$ is valid.

**II. Perturbation theory**

To obtain the spatial distribution of the polarization at small positive biases, $V$, Eq. (A.1a) was linearized as $P_3(\mathbf{r}) = -P_0(x) + p(\mathbf{r})$, where $P_0(x)$ is the initial flat domain wall profile positioned at $x = x_0$:

$$P_0(x) = P_S \tanh\left((x - x_0)/2L_\perp\right). \tag{A.5}$$

where the correlation length is $L_\perp = \sqrt{-\eta/2\alpha}$, and the spontaneous polarization is $P_S^2 = -\alpha/\beta$.

Polarization $p(\mathbf{r})$ is the induced due to materials response to a biased probe. The condition $p(\mathbf{r}) \to 0$ is valid far from the probe at an arbitrary applied bias. Here, we derive the solution within a perturbation approach.

Under the condition of a thick sample, $h \gg d$, the approximate closed form expression for the linearized stationary solution of Eq. (A.2) is derived as (see Supplement in Ref.[40] for details):

$$p(\rho, z) \approx \frac{V^*}{(d + L_\perp)\alpha_S} \left( \frac{(d + z/\gamma)d^2}{\left(L_\perp(d + z/\gamma) + (d + z/\gamma)^2 + \rho^2\right)^{3/2}} + \frac{d^2(d^2 + \rho^2) - 3d^4}{\gamma(d^2 + \rho^2)^{5/2}} L_z \exp\left(-\frac{z}{L_z}\right) \right). \tag{A.6}$$

Here $\rho = \sqrt{x^2 + y^2}$ is the radial coordinate. The correlation length $L_z = \sqrt{\varepsilon_0 \varepsilon_{33}^b \xi}$ is extremely small for typical values of gradient term $\xi$. The effective dielectric anisotropy factor $\gamma = \sqrt{\gamma_b^2 + 1/(\varepsilon_{11}\varepsilon_0 \alpha_S)}$ and the "bare" dielectric anisotropy factor $\gamma_b = \sqrt{\varepsilon_{33}^b/\varepsilon_{11}}$ are introduced.

When deriving expression (A.6), we utilized the inequalities $2\varepsilon_0 \varepsilon_{33}^b |\alpha_S| \ll 1$, $\varepsilon_{33}^b \ll \varepsilon_{33}$, $L_\perp \leq 0.5...5$ nm and $L_z < 1$ Å, valid for typical ferroelectric material parameter $\xi \sim 10^{-8}...10^{-10}$ J m$^3$/C$^2$ and the background permittivity $\varepsilon_{33}^b \leq 5$. Assuming the validity of additional inequalities $L_z \ll L_\perp \ll d$, the approximate solution was derived as:

$$p(\rho, z) \approx \frac{V}{\alpha_S d} \frac{(d + z/\gamma) d^2}{\left((d + z/\gamma)^2 + \rho^2\right)^{3/2}}, \quad \text{at} \quad z \gg L_z, \tag{A.7a}$$

$$E_3(\rho, z) \approx \frac{V(d + z/\gamma) d}{\gamma\left((d + z/\gamma)^2 + \rho^2\right)^{3/2}}. \tag{A.7b}$$

The linear approximation for the polarization distribution given by Eq.(A.7) is quantitatively valid until $|p| < P_S$ or, alternatively, $V/d < \alpha P_S$, i.e. at biases $V$ much smaller than the coercive bias, at which polarization reversal is absent. This means that the probe induced domain formation cannot be considered quantitatively within the linearized LGD-equation.

Below we take into account the ferroelectric material nonlinearity within direct variation method. Using trial function with variational parameter $P_V$

$$P_3(x, y, 0) \approx P_0(x) + \frac{\sqrt{\varepsilon_{11}\varepsilon_0/(-2\alpha)}\, d^2 \cdot P_V(V)}{\left(d\sqrt{\eta/(-2\alpha)} + d^2 + x^2 + y^2\right)\sqrt{d^2 + x^2 + y^2}}. \tag{A.8}$$

one can obtain renormalized free energy.

Under the reasonable assumption $L_\perp \ll d$, polarization distribution (A.8) produces the following depolarization field:

$$E_3^d(\rho, z) \approx \frac{\gamma \gamma_b^2}{\gamma^2 - \gamma_b^2} \frac{d^2 P_V}{\varepsilon_0 \varepsilon_{33}^b} \left( \frac{(d + z/\gamma)}{\gamma \left((d + z/\gamma)^2 + \rho^2\right)^{3/2}} - \frac{(d + z/\gamma_b)}{\gamma_b \left((d + z/\gamma_b)^2 + \rho^2\right)^{3/2}} \right). \quad (A.9)$$

Polarization distribution is shown in Fig. A.1.

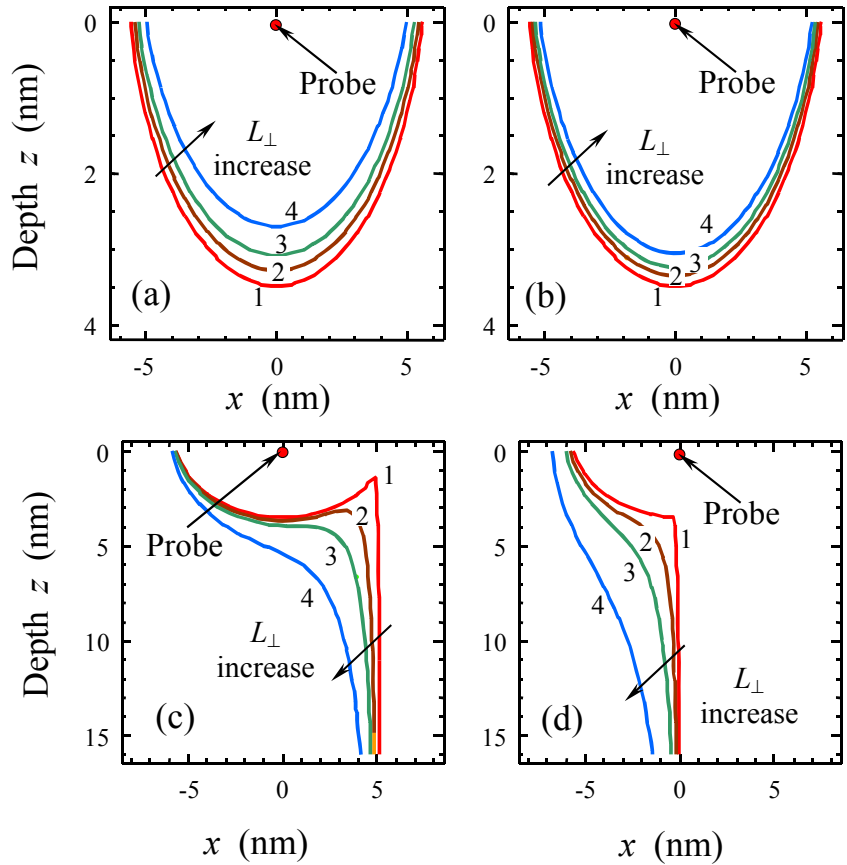

**Fig. A.1.** Domain wall vertical cross-section for different distances from initial flat wall $x_0 = \infty$, 15, 5, 0 nm (panels (a), (b), (c) and (d) respectively). Curves 1, 2, 3, 4 corresponds to different values of $L_\perp = 0$, 0.5, 1, 2 nm. Other parameters: effective distance $d = 5$ nm, $\varepsilon_{11} = 500$, $\alpha = -1.66 \cdot 10^8$ m/F, $\beta = -1.44 \cdot 10^8$ m$^5$/(C$^2$F).

After the substitution of Eq.(A.8-9) into the free energy functional (S.1a) and integration, renormalized free energy was derived as

$$\Delta\Phi(P_V) = \pi d \sqrt{\frac{\varepsilon_0 \varepsilon_{11}}{-2\alpha}} \left( -P_V V + \frac{w_1}{2} P_V^2 + \frac{w_2}{3} P_V^3 + \frac{w_3}{4} P_V^4 \right), \qquad (A.10a)$$

$$w_1 \approx 1, \quad w_2(x_0) \approx \frac{-3\beta P_S x_0}{\sqrt{(L_\perp + d)^2 + x_0^2}} \frac{\sqrt{-2\alpha\varepsilon_{11}\varepsilon_0}}{4\alpha^2(L_\perp + d)}, \quad w_3 \approx \frac{\beta\varepsilon_{11}\varepsilon_0}{4\alpha^2(L_\perp + d)^2} \qquad (A.10b)$$

it is easy to find the equation of state from Eq.(S.8a). So, in the presence of lattice pinning of viscous friction type, the amplitude $P_V$ should be found from Landau-Khalatnikov equations as:

$$\tau \frac{d}{dt} P_V + w_1 P_V + w_2(x_0) P_V^2 + w_3 P_V^3 = V(t) \qquad (A.11)$$

The parameter $P_V$ serves as effective variational parameter describing domain geometry, and allows reducing complex problem of domain dynamics in the non-uniform field to an algebraic equation Eq. (A.11).

Critical points of polarization bias dependence (inflection points, coercive biases) could be found from the static equation $dV/dP_V = 0$, namely we derived expressions for coercive biases $V_c^\pm$, loop halfwidth $\Delta V_c$ and imprint bias $V_I$ as

$$V_c^\pm = \frac{w_2(2w_2^2 - 9w_3 w_1) \pm 2(w_2^2 - 3w_3 w_1)^{3/2}}{27 w_3^2},$$

$$\Delta V_c = \frac{V_c^+ + V_c^-}{2} = \frac{2(w_2^2 - 3w_3 w_1)^{3/2}}{27 w_3^2}, \quad V_I = \frac{V_c^+ - V_c^-}{2} = \frac{w_2(2w_2^2 - 9w_3 w_1)}{27 w_3^2}. \qquad (A.12)$$

### III. Effective piezoresponse calculations

In decoupled approximation and object transfer functions approach (see Refs.[41, 42, 43]), analytical $V$ dependence of effective piezoelectric response $PR(V)$ were found as:

$$PR(V, x_0) = d_0^{eff}(x_0) - \frac{\varepsilon_0 \varepsilon_{11}}{\gamma} \sum_{i=1}^{4} \frac{B_i(\gamma) \cdot \ln(e + b_i(\gamma)/C) P_V(V)}{(b_i(\gamma) + \ln(e + b_i(\gamma)/C))(L_\perp + d \ln(e + b_i(\gamma)/C))}. \quad (A.13)$$

Where $d_0^{eff}(x_0)$ is the bias-independent PFM profile of the flat 180°-domain wall located at distance $x$ from the tip apex. Response $d_0^{eff}(x_0)$ was calculated in Ref.[44]; dielectric anisotropy factor is $\gamma = \sqrt{\varepsilon_{33}/\varepsilon_{11}}$. Constants $e \approx 2.71828...$ is the natural logarithm base and $C \approx 0.577216...$ is Euler's constant. Constants $b_1(\gamma) = \gamma^2/(1+\gamma)^2$, $b_2(\gamma) = \gamma/(1+\gamma)$, $b_3(\gamma) = \gamma(1+2\gamma)/2(1+\gamma)^2$, $b_4(\gamma) = \gamma(16-15\gamma^2)/4(1+\gamma)^2$ and $B_1 = -2\varepsilon_0 \varepsilon_{33} Q_{12} \gamma/(1+\gamma)^2$, $B_2 = 2(1+2\nu)\varepsilon_0 \varepsilon_{33} Q_{12}/(1+\gamma)$, $B_3 = 2\varepsilon_0 \varepsilon_{33} Q_{11}(1+2\gamma)/(1+\gamma)^2$, $B_4 = 2\varepsilon_0 \varepsilon_{11} Q_{44} \gamma^2/(1+\gamma)^2$ ($\nu$ is Poisson ratio, $Q_{ij}$ is electrostriction tensor for cubic symmetry).

The expected behavior of the hysteresis loops as a function of tip surface separation is illustrated in Fig.A.2. Directly at the wall, the loop is closed and the local response originates from the bias-induced bending of the domain wall. It is clear from the figure, that the loop halfwidth, determined as the difference of coercive biases $\Delta V_c = (V_c^+ - V_c^-)/2$, appears and monotonically increases with the distance $x$ increase. The bistability is possible and $\Delta V_c$ is defined only for $x_0^2 \geq 2(L_\perp + d)^2$. Far from wall ($x_0 >> d$) corresponding coercive biases are symmetric, $V_c^\infty = 2\alpha P_S(L_\perp + d)/3\sqrt{-6\alpha\varepsilon_{11}\varepsilon_0}$. The inclusion of viscous friction leads to the loop broadening and smearing far from the wall, while near the wall the minor loop opening is observed (compare solid and dotted curves). Note that the observed evolution of the loop shape and switching parameters agrees with the experimental observations.

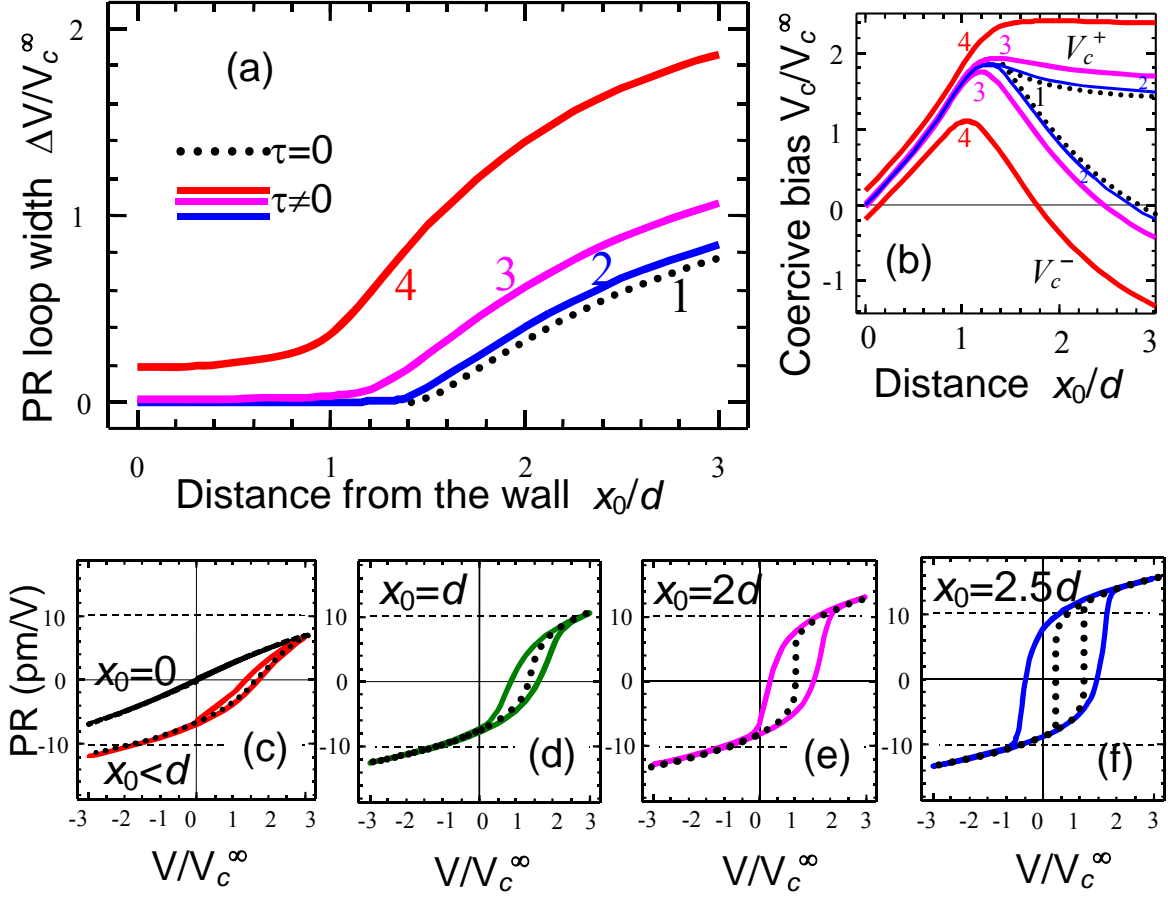

**Fig. A2.** (a) Piezoresponse (PR) loop relative width $\Delta V_c / V_c^\infty$ ($V_c^\infty$ is the static coercive bias far from the wall) vs. the distance from the wall $x_0/d$. (b) Left (bottom curves) and right (top curves) coercive biases of the PR loops. Curves 1, 2, 3, 4 correspond to the different relaxation coefficients $\tau$=0, $10^{-8}$, $10^{-7}$, $10^{-6}$ SI units. Plots (c-f) show piezoresponse loops vs. applied bias (V) calculated for increasing distance $x_0$ from domain wall (labels at the plots). Material parameters for $LiNbO_3$ are $\varepsilon_{11}$=84, $\alpha = -1.95 \cdot 10^9$ m/F, $\beta = 3.61 \cdot 10^9$ m$^5$/(C$^2$F),

$P_S$=0.75 C/m$^2$, Poisson ratio is $\nu$=0.3, parameter $d$=60 nm, frequency $\omega=2\pi\cdot 10^4$ rad sec$^{-1}$ and maximal bias $U_{max}$=15 V.

**IV. The influence of the probe tip conical part on the domain nucleation**

The tip of the probe induces strong but localized electric field, while the conical part of the probe produces weaker but more diffused field distribution. The influence of the probe tip conical part on the domain nucleation is shown in Fig. A3. This effect is evident from Figs. A3a,b, since the nascent domain is more diffuse for the case with conical part included. It is also seen from Figs. A3c, d that the flat domain wall is practically unaffected by the field of probe tip for distances $x_0$>10 $d$ between them, while the conical part field induces wall bending even in this region.

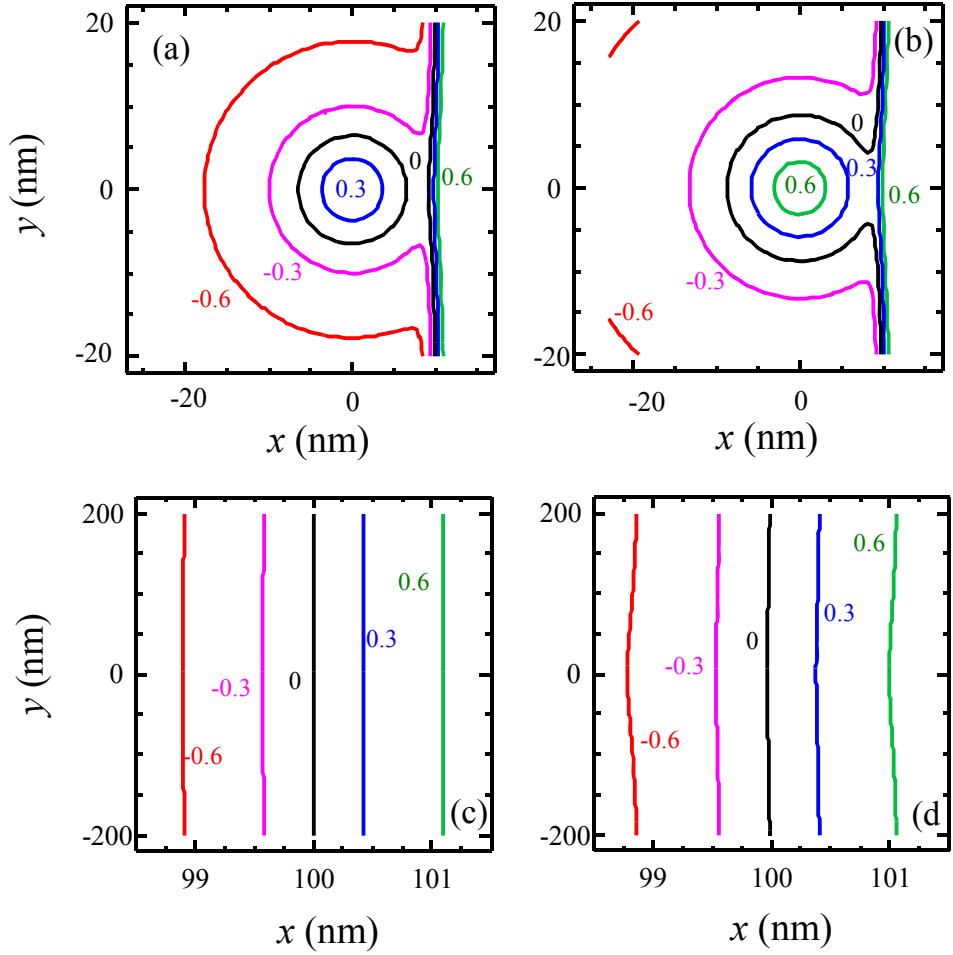

**Fig. A3**. Contour maps of the bound charge distribution (values near the curves) on the surface ($z=0$) for nucleation near the flat wall at $x_0=10$ nm (a, b) and far from the flat wall at $x_0=100$ nm (c, d); for two different probe models, effective point charge alone (a, c) and effective point and line charges (b, d). For the tip far from the wall only the near wall region is shown. Effective distance between the charge and surface $d=10$ nm, applied voltage $V=30$ V, line charge length is 1 μm.

Thus, the long-range influence of the probe conical part on the initial domain wall behavior could explain logarithmically slow saturation of nucleation bias shown in Fig.A4.

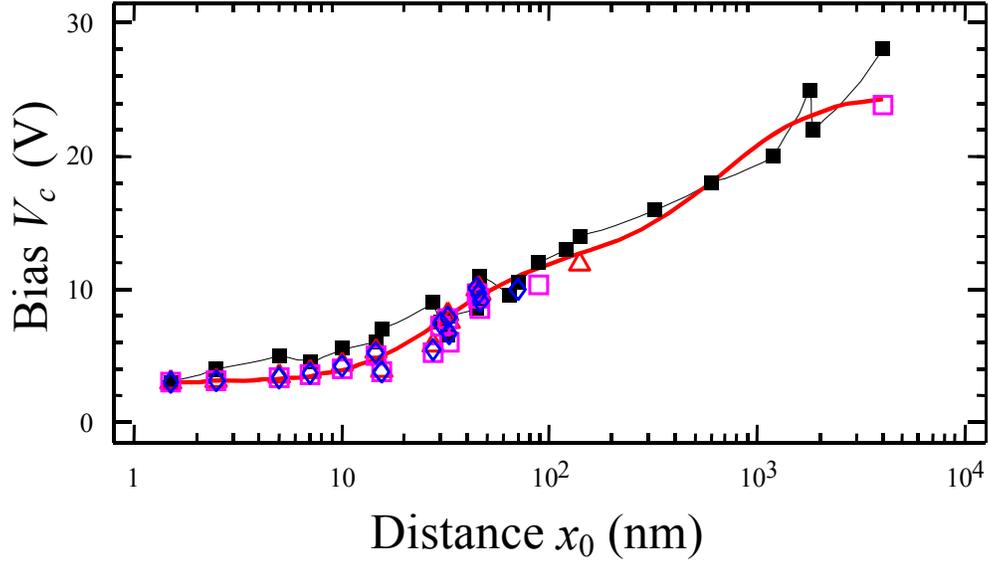

**Fig.A4.** (a) PFM hysteresis loop halfwidth $\Delta V_c = (V_c^+ - V_c^-)/2$ vs. the distance from the wall. Material parameters for LiNbO$_3$ are $\varepsilon_{11}$=84, $\alpha$= $-$ 1.95·10$^9$ SI units, $P_S$=0.75 C/m$^2$, Poisson ratio is $\nu$=0.3; domain wall intrinsic width $L_\perp$=0.5 nm. Filled boxes are experimental points. Red solid curve is the fitting for the model that takes into account point charge with $d$=30 nm and line charge $L$ spanning from 100 nm to 1000 nm. Blue dashed curve is the fitting with the equation $\Delta V_c = 7.6 \cdot \lg(2.2 + x_0/2.6)$ (with x$_0$ in nm). Theoretical curves are calculated for threshold bias $V_{th}$ = 3 V originated from the lattice pinning.

**Appendix B. Calculations of nucleation bias within MW and BC approaches**

Excess free energy for nascent nucleus at the domain under the external field after Miller and Weinreich [21] is

$$F = -2P_S \langle W_0 \rangle + 2\sigma_W c \sqrt{a^2 + l^2} + 2\sigma_p c \frac{a^2}{l} \qquad (B.1)$$

First term is the energy of nucleus interaction with external field, second term is the excess wall energy and the third one is the depolarization field energy. Here $P_S$ is the spontaneous polarization, $\langle W_0 \rangle$ is the external field $E_0$ and integrated on the nucleus volume, $c$ is the nucleus width normal to the wall, $a$ is the nucleus half-width on the surface along the wall, $l$ is the size of nucleus along the wall and normal to the surface (see Fig. B1). The surface energy of the domain wall $\sigma_W$ is regarded independent on the wall orientation. $\sigma_p$ is an effective surface density of the depolarization field energy, $\sigma_p \approx \ln(0.74 a/c) P_S^2 c / (\pi \varepsilon_0 \varepsilon_{11})$ in SI units. Here width $c$ was regarded of lattice constant order and considered much smaller than other sizes of nucleus.

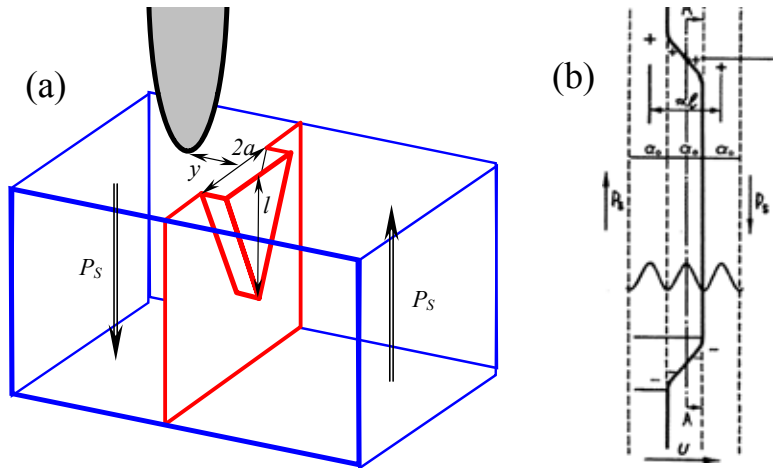

**Fig. B1.** Schematics of calculations: (a) triangular prism nucleus from Miller and Weinreich, (b) Burtsev-Chervonobrodov smooth nucleus.

For the case of homogeneous external field, considered by Miller and Weinreich, $\langle W_0 \rangle$ is simply the product of nucleus volume and the electric field value $\langle W_0 \rangle \approx E_0 c_0 a l$. For considered case of inhomogeneous electric field of SPM probe $\langle W_0 \rangle$ is

$$\langle W_0 \rangle = \begin{pmatrix} \dfrac{2c\,aVd}{\gamma\sqrt{a^2+(l/\gamma)^2}} \ln\left( \dfrac{a^2 - d\,l/\gamma + \sqrt{(a^2+(l/\gamma)^2)(a^2+d^2+y^2)}}{-(d+l/\gamma)l/\gamma + \sqrt{(a^2+(l/\gamma)^2)((d+l/\gamma)^2+y^2)}} \right) \\ -\dfrac{2c\,Vd}{\gamma} \ln\left( \dfrac{a+\sqrt{a^2+d^2+y^2}}{\sqrt{d^2+y^2}} \right) \end{pmatrix} \quad \text{(B.2)}$$

Here $V$ is the bias, applied to the probe, $d$ is the effective charge –surface distance, $y$ is the distance between the probe axis and the domain wall, $\gamma = \sqrt{\varepsilon_{33}/\varepsilon_{11}}$ is the dielectric anisotropy factor.

When the nucleus sizes are small $\langle W_0 \rangle \approx \dfrac{d^2 V c a l}{\gamma\left(\sqrt{d^2+y^2}\right)^3}$ and

$F = -2P_S \dfrac{d^2 V c a l}{\gamma\left(\sqrt{d^2+y^2}\right)^3} + 2\langle\sigma_W\rangle c\sqrt{a^2+l^2} + 2\sigma_p c\dfrac{a^2}{l}$. It is seen, that in this case the free energy $F$ in the inhomogeneous field is the same as in homogeneous one, but with substitution of $E_0$ with $E_P(V) = \dfrac{d^2 V}{\gamma\left(\sqrt{d^2+y^2}\right)^3}$. Thus, Miller-Weinreich activation energy of domain wall step nucleation, obtained with respect to probe tip electric field inhomogeneity, is

$$F_a(\sigma_W, V, x_0) = \dfrac{8}{3\sqrt{3}} \sqrt{\ln\left( \dfrac{\sigma_W \gamma\left(\sqrt{d^2+x_0^2}\right)^3}{2cP_S d^2 V} \right) \dfrac{(c\sigma_W)^3}{\pi\varepsilon_0\varepsilon_{11}} \dfrac{\gamma\left(\sqrt{d^2+x_0^2}\right)^3}{d^2 V}}. \quad \text{(B.3)}$$

Directly at the wall ($x_0=0$)

$$F_a^{MW}(\sigma_W, V) = \frac{16}{3\sqrt{3}} \sqrt{\ln\left(\frac{\sigma_W \gamma d}{2cP_S V}\right) \frac{(c\sigma_W)^3}{4\pi\varepsilon_0 \varepsilon_{11}} \frac{\gamma d}{V}}. \qquad (B.4)$$

It should be noted, that Miller and Weinreich considered lattice discreteness in very rough model and do not take the possibility of wall to bent into account. Burtsev and Chervonobrodov considered a more realistic model with continuous lattice potential and diffuse domain walls, at that the nucleus shape and domain wall width are calculated self-consistently. Using their approach we obtained expression:

$$F_a^{BC}(\sigma_W, V) = \sqrt{\ln\left(\frac{\gamma d \sqrt{\sigma_{min} \delta\sigma}}{2cP_S V}\right) \frac{\left(c\sqrt{\sigma_{min} \delta\sigma}\right)^3}{4\pi\varepsilon_0 \varepsilon_{11}} \frac{\gamma d}{V}}. \qquad (B.5)$$

Using dependence of activation energy on applied bias, one could find activation voltage from the equality of activation energies (B.4-5) to some relevant level.

Below we used the following values of lattice potential: minimal value $\sigma_{min} = 0.160 \, J/m^2$ and modulation depth $\delta\sigma_{min} = 0.150 \, J/m^2$ calculated for domain walls in LiNbO$_3$. Other parameters were $c$=0.5 nm, $P_S$=0.75 C/m$^2$, $\varepsilon_{11} = 84$ $\varepsilon_{33} = 30$, effective distance $d$ (tip size) was determined from the expression $d = V_c \sqrt{\frac{27\beta\varepsilon_{11}\varepsilon_0}{2\alpha^2}}$, where $V_c$ is coercive bias far from the wall.

In Table 1 we presented results of activation voltage calculations for the models of Miller-Weinreich and modified Burtsev-Chervonobrodov.

**Table 1.** Values of activation voltage for domains wall in LiNbO$_3$

| Model | Barrier level $F_a(\sigma_W, V, x_0 = 0)$ |
|---|---|
|  |  |

|    | $25\,k_B T$ | | | $k_B T$ | | |
|----|---|---|---|---|---|---|
|    | $d$ values (nm) | | | $d$ values (nm) | | |
|    | 21 | 61 | 86 | 21 | 61 | 86 |
| MW | 3.9 V | 11.4 V | 16.1 V | 5.2 V | 15.1 V | 21.2 V |
| BC | 0.9 V | 2.6 V | 3.6 V | 2.6 V | 7.4 V | 10.5 V |